\documentclass[10pt,conference]{IEEEtran}
\usepackage{cite}
\usepackage{amsmath,amssymb,amsfonts}
\usepackage{algorithmic}
\usepackage{graphicx}
\usepackage{textcomp}
\usepackage[disable]{todonotes}
\usepackage{xcolor}
\def\BibTeX{{\rm B\kern-.05em{\sc i\kern-.025em b}\kern-.08em
    T\kern-.1667em\lower.7ex\hbox{E}\kern-.125emX}}
\begin{document}

\title{Shortest Path in Pauli Forest - An Algorithm for Decomposing Pauli Exponentials to Quantum Circuits
}

\author{\IEEEauthorblockN{1\textsuperscript{st} Lauri Vuorenkoski}
\IEEEauthorblockA{\textit{Department of Computer Science} \\
\textit{University of Helsinki}\\
Helsinki, Finland \\
lauri.vuorenkoski@helsinki.fi}
\and
\IEEEauthorblockN{2\textsuperscript{nd} Arianne Meijer--van de Griend}
\IEEEauthorblockA{\textit{Department of Computer Science} \\
\textit{University of Helsinki}\\
Helsinki, Finland \\
arianne.vandegriend@helsinki.fi}
}

\maketitle

\begin{abstract}
Decomposing Pauli exponentials efficiently to quantum circuits has been the subject of intense research in recent years. Pauli exponentials are an essential component of many different quantum algorithms. Due to the error-prone nature of current and near term quantum devices, it is crucial that quantum circuits are as compact as possible. Several different types of algorithms have been developed to decompose Pauli exponentials into as short circuits as possible. We propose a novel algorithm for architecture-aware synthesis of Pauli exponentials that also determines the initial qubit placement on the device. We call this the Shortest Path in Pauli Forest algorithm. The results show an improved CNOT count and runtime for both random Pauli exponentials and molecular ans\"atze.
\end{abstract}


\section{Introduction}

In NISQ (Noisy Intermediate Scale Quantum)~\cite{preskill_quantum_2018} computing, it is essential that quantum circuits be as compact as possible. Due to short decoherence times and noise, the error in the execution of the circuit accumulates as the length of the circuit increases~\cite{schlosshauer_decoherence_2005}. One key factor is the number of CNOTs and the CNOT depth since CNOT gates are noisier and take longer than single-qubit gates~\cite{perez_error-divisible_2023}.

Another crucial limitation of some NISQ era devices is qubit connectivity~\cite{li_tackling_2019}. Typically, two qubit gates can only be applied on certain qubit-pairs on the physical device. When the quantum algorithm is transpiled to a circuit that can be run on a physical device, it must be accommodating to limitations of qubit connectivity. The simplest way to overcome this is to include SWAP gates. However, this significantly increases CNOT count. In this paper, we describe a novel algorithm for efficiently decomposing Pauli exponentials into quantum circuits. 

The structure of the paper is as follows: In the second section, we briefly describe Pauli exponentials, Pauli gadgets, the used quantum gates, and Clifford tableaux followed by describing existing algorithms to decompose Pauli exponentials into quantum circuits. In the third section, we describe the idea of the novel algorithm. In the fourth section, we benchmark our algorithm against previous algorithms with both random Pauli exponentials and those representing molecular hamiltonians. In the final section, we discuss some of the key findings of the study.

\section{Preliminaries and Related Work}
This section gives a short overview of the necessary theory behind our architecture-aware synthesis, as well as an overview of related algorithms that do something similar. First, the definition of a Pauli exponential is given, followed by that of the Clifford tableau formalism. The section is concluded with an overview of previous work.

\subsection{Pauli Exponentials}

A \textit{Pauli exponential} is a sequence of multi qubit rotations, $\prod e^{-i \alpha I\otimes X\otimes Y \otimes Z}$, where each term in the product is called a \textit{Pauli gadget}~\cite{cowtan_phase_2020} and each tensor product $I\otimes X\otimes Y \otimes Z$, called the \textit{Paulistring}, is often abbreviated as $IXYZ$. The Paulistring defines the axis of rotation for each qubit involved in the interaction, and $\alpha$ is the angle of rotation.

Mutually commuting Pauli gadgets in a Pauli exponential can be arbitrarily reordered. This is not always the case, but in some special cases, we can reorder anti-commuting gadgets. For trotterized time evolution operators in particular, reordering the Pauli gadgets changes the already existing Trotter error. In this case, the Pauli exponential that approximates the time evolution is replaced by a different approximation that has a similar Trotter error, but less gate error due to the improved CNOT count. This is a popular strategy for Hamiltonian synthesis~\cite{brugiere_faster_2024, huang_redefining_2024, li_paulihedral_2022, schmitz_graph_2023, cowtan_generic_2020}. 

\subsection{Quantum Circuit}

In this paper, we use $V=\sqrt{X}$, $S=\sqrt{Z}$, $V^\dagger$, $S^\dagger$, $CNOT$ and $R_z$ gates to decompose Pauli exponentials into quantum circuits.

\begin{equation}
V= 1/2 \begin{pmatrix}
1+i & 1-i \\
1-i & 1+i
\end{pmatrix} \quad
S= \begin{pmatrix}
1 & 0 \\
0 & i
\end{pmatrix}  
\end{equation}

When the $V^\dagger$ gate is commuted through the Pauli operator, the $X$ component of the Pauli operator is added to the $Z$ component (modulo 2) so that it changes operator $Z$ to $Y$ and vice versa. The gate commutes with operator $X$. The $S^\dagger$ gate has a similar effect for the $X$ component. It adds the $Z$ component to the $X$ component such that it changes operator $X$ to $Y$ and vice versa. The gate commutes with operator $Z$.

Commuting the $S^\dagger$ gate first, followed by the $V^\dagger$ gate has the effect that $Z$ is changed to $Y$, $Y$ is changed to $X$, and $X$ is changed to $Z$. Applying these two gates in the reverse order has the reverse effect. 

Commuting a $CNOT$ gate through the Pauli operator on the control qubit $c$ and the Pauli operator on the target qubit $t$ adds the $X$ component of operator $c$ to the $X$ component of operator $t$ and the $Z$ component of operator $t$ to the $Z$ component of operator $c$.

\subsection{Clifford Tableau}

Our algorithm utilizes Clifford tableau synthesis~\cite{aaronson_improved_2004} for making the last part of the resulting circuit more compact. The Clifford tableau represents a quantum circuit composed of Clifford gates in a compact binary table of size $2n \times 2n$ and a binary vector of size $2n$. The tableau of an empty circuit is an identity matrix. Appending and prepending Clifford gates to a Clifford operator represented by the Clifford tableau corresponds to row and column additions in the Clifford tableau~\cite{huang_redefining_2024}. 

A Clifford tableau can be synthesized to a quantum circuit using several methods. In this paper, we used the same algorithm as Huang et al.~\cite{huang_redefining_2024}. Usually, the method produces a circuit with significantly fewer quantum gates compared to the original circuit~\cite{winderl_architecture-aware_2024}.

\subsection{Existing Algorithms}

A simple method to decompose a Pauli gadget into a quantum circuit is to first diagonalize the Pauli gadget by changing all Pauli operators to $Z$. This is achieved by conjugating the gadget with $S$ and $SV$ gates and their adjoints. After this, all the qubits except one are disconnected from the Pauli gadget by conjugating the gadget with a CNOT ladder. Lastly, the $R_Z$ rotation is applied to the last remaining qubit. However, during this process, there is a lot of freedom in choosing which CNOTs to apply and making the wrong choices can make it so that the CNOTs between subsequent Pauli gadgets are not canceled out even though they could have been.

This is very inefficient in terms of CNOT count and CNOT depth. To overcome this, several more efficient algorithms have been designed to do this. 

The set-based algorithm of Cowtan et al.~\cite{cowtan_generic_2020} partitions Pauli gadgets into mutually commuting sets (a similar idea is presented by van de Berg et al.~\cite{berg_circuit_2020}). The main objective of the algorithm is to reduce the 2-qubit gate depth. Pauli gadgets are grouped into a minimum number of commuting sets. The gadgets are processed set by set. First, all gadgets of a set are diagonalized to phase gadgets, which is always possible when Pauli gadgets commute with each other. After that, phase gadgets are processed to a circuit using GraySynth~\cite{amy_controlled-not_2018}. The algorithm is included in TKET and it is not architecture aware, so it assumes that we can place CNOTs between any arbitrary pair of qubits. 

The Paulihedral algorithm~\cite{li_paulihedral_2022} predefines some Pauli gadgets to be processed together in blocks where all Pauli strings share the same parameter. The Pauli strings can be reordered within each block. The algorithm utilizes CNOT gate cancellations of neighboring synthesized gadgets and decomposes each Pauli gadget such that it avoids additional swaps when taking the target architecture into account. Blocks can be ordered lexicographically to increase gate cancellations.

The algorithm based on the Pauli Frame Graph is developed by Schmitz et al.~\cite{schmitz_graph_2023} and implemented in TKET. It utilizes gadget reordering and theoretically views the Pauli exponential synthesis problem as a graph problem, where the nodes represents distinct Pauli frames (similar to the Clifford tableau) and edges represent nine different 2-qubit entangling gates altering the Clifford tableau (composed of single qubit gates and CNOT gates). 

The problem is then equivalent to solving a problem similar to the travelling salesman problem: what is the optimal path to visit all nodes in which all gadgets can be decomposed. A gadget can be decomposed into several nodes, and it is possible that several gadgets can be composed into the same node. 

The path is constructed in a greedy fashion. At every step, the algorithm selects the smallest gadgets. Then, it identifies the 2-qubit entangling gate such that it will bring at least one of the small gadgets smaller. From these options, the algorithm selects one gadget depending on how it brings the other gadgets smaller and how it affects the depth of the circuit. This algorithm is not architecture-aware.

The objective of the Rustiq algorithm is to minimize the CNOT count or the CNOT depth~\cite{brugiere_faster_2024}. The algorithm assumes that all gadgets are commuting, reordering the gadgets as it sees fit. Gates are added to the circuit one by one so that Pauli operators of the next gadget are turned into I. The gates used are CNOTs (applied in either direction), $I$, $H$ or $V$ for the control qubit and $I$, $H$, $S$, for the target qubit (all together 18 options). Each combination of gates for a qubit-pair is selected such that maximizes the amount of non-$I$ Pauli letters in the Pauli exponential that are turned to $I$ in all remaining Pauli gadgets. The algorithm is not architecture-aware.

Huang et al.~\cite{huang_redefining_2024} proposed an alternative that is architecture-aware by generalizing the well-known GraySynth~\cite{amy_controlled-not_2018} algorithm to Pauli exponentials. Additionally, the algorithm uses Clifford tableau synthesis to reduce the CNOT count in end of the circuit where all gates are Clifford gates. The algorithm was benchmarked against Paulihedral and TKET with random Pauli exponentials and Pauli exponentials representing different molecular ans\"atze. The algorithm produced circuits with significantly lower CNOT counts and shallower CNOT depth than Paulihedral or TKET.

The Pauli Forest algorithm maps gadgets as trees trying to minimize CNOT depth~\cite{li_pauliforest_2025}. Each gadget is mapped to a unique tree such that the number of nodes is minimized. Architecture constraints are followed when building the tree, so this algorithm is architecture-aware. The algorithm also includes an initial mapping part, which is based on mapping closely related logical qubits near each other into a physical connectivity map. The algorithm does not reorder the gadgets during synthesis. 

\section{Methods}

In this work, we model the problem as a graph where the objective is to find the shortest path that ensures all Pauli gadgets are decomposed upon completion. In this formulation, the edges of the graph represent Clifford gates, while the nodes correspond to unique Clifford tableaus representing the cumulative sequence of gates applied from the initial state. Consequently, the starting node is defined by the identity tableau.

Each Pauli gadget may be decomposed at multiple nodes in the graph, reflecting the various strategies available for decomposition. Furthermore, a single node can facilitate the simultaneous decomposition of multiple Pauli gadgets, provided those gadgets share a high degree of similarity.

This approach builds upon the Pauli Frame Graph algorithm ~\cite{schmitz_graph_2023}. However, that original algorithm is not architecture-aware and does not incorporate Clifford tableau synthesis. They conceptualized the task as a Travelling Purchaser Problem (related to NP-hard Travelling Salesman Problem) in which a shopper has a list of items to buy. Items are sold in several specific stores, and the same store can sell several items. The shopper must travel to stores to buy every item on the list using the shortest possible route.

Using this idea, we propose a new synthesis and mapping algorithm for Pauli exponential synthesis which we call Shortest Path in Pauli Forest. The algorithm produces circuits with a distinct structure. Namely, each output circuit starts with gates from the Clifford+$R_Z$ gate set and ends with a long sequence of (trailing) Clifford gates. The trailing Clifford gates are the adjoint of the Clifford gates in the first part, but with the $R_Z$ gates omitted. Since the second part typically contains many gates, it can be optimized by storing the transformation in a Clifford tableau and synthesizing a smaller circuit instead.

\begin{figure}
\centerline{\includegraphics[width=0.45\textwidth]{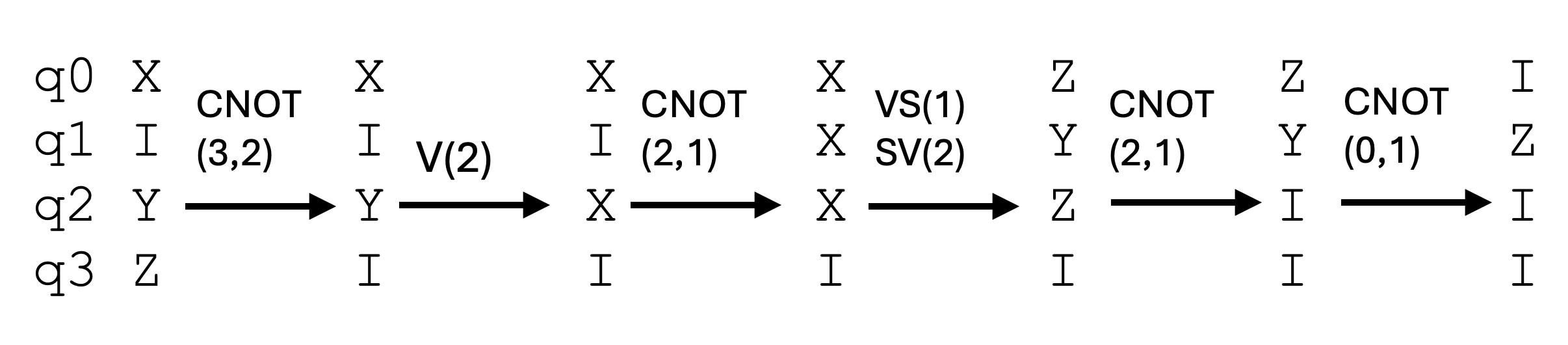}}
\caption{Example of the decomposition of a single Pauli gadget restricted to a line topology. Distance of the path in this example is 4.}
\label{fig:decomposition}
\end{figure}

The algorithm decomposes Pauli gadgets one by one (see Fig.~\ref{fig:decomposition} for example decomposition). When synthesizing a Clifford gate, it is appended to the first part, and its adjoint is prepended to the second part. All remaining Pauli gadgets are transformed when the adjoint gate is commuted through the gadget. When a gadget only has a single remaining non-$I$ operator left, it is diagonalized and a corresponding $R_Z(\theta)$ gate is appended to the first part. 

The steps needed to decompose one Pauli gadget can be seen as a path consisting of single qubit gates and CNOT gates. Single qubit gates are used for diagonalization and CNOT gates for disconnecting qubits in a Pauli gadget. 

The distance of a path is calculated by the minimum number of CNOTs needed. Synthesizing a Pauli gadget requires at least $n-1$ CNOT gates when it contains $n$ non-identity Pauli operators ($X$, $Y$, $Z$). Due to connectivity constraints, we need to use bridge qubits when qubits participating in the Pauli gadget are not connected. In these cases, we first need to add these bridge qubits to the Pauli gadget. This is done using an additional CNOT gate ($q1$ in the Fig.~\ref{fig:decomposition}).

The Pauli gadget can be decomposed using many different paths. One of the choices is which qubit will hold the last non-$I$ Pauli operator that is then decomposed using single qubit Clifford gates and an $R_Z$ gate. There are also many other choices in Pauli gadget decomposition: for example, which single qubit gates are used and how the possible bridge qubits are added to the gadget.

All gates which are prepended to the second part of the circuit must be propagated through remaining Pauli gadgets which can change the remaining gadgets. This can shorten or lengthen the distance of the path to decompose remaining Pauli gadgets after propagation.

The algorithm assumes, without loss of generality, that Pauli gadgets are mutually commuting. In the case of circuits representing a trotterized Hamiltonion, this allows us to exchange gate error for Trotter error. Thus, the algorithm exploits the fact that we can decompose the Pauli gadgets in arbitrary order. The next gadget to be processed is greedily selected according to the distance metric, similar to the Rustiq and Pauli Frame Graph algorithms~\cite{brugiere_faster_2024, schmitz_graph_2023}. The gadget with the shortest distance is processed next. 

\subsection{Algorithm for Initial Mapping}

\begin{figure}
\centerline{\includegraphics[width=0.35\textwidth]{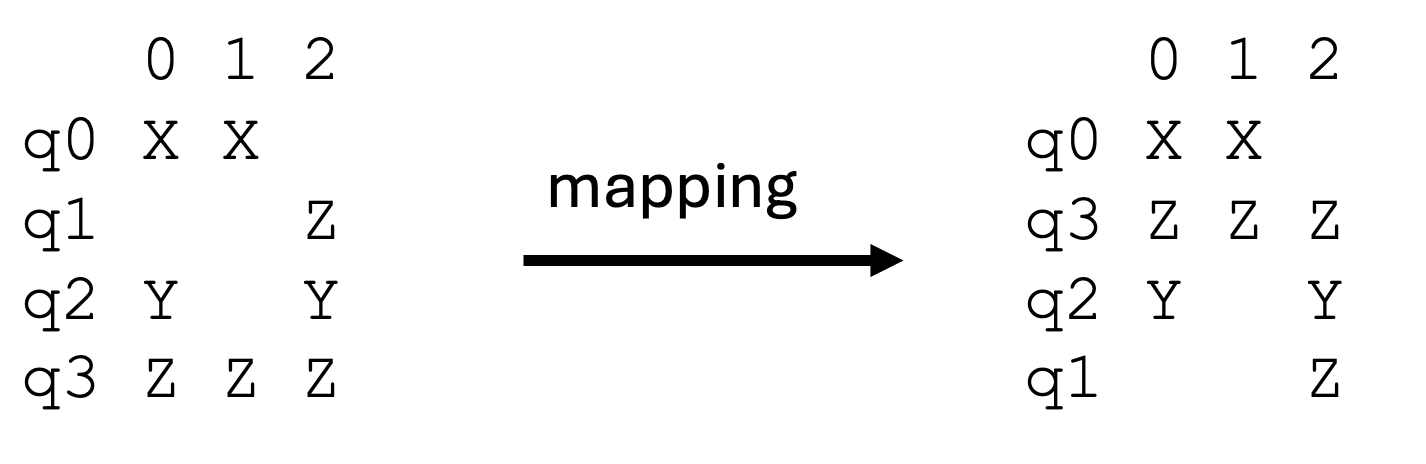}}
\caption{Example of optimal mapping with line topology 0-1-2-3.}
\label{fig:mapping}
\end{figure}

The idea of the initial mapping algorithm is to shorten the distance from the starting point to the points that decompose the Pauli gadgets. This is done by trying to minimize the needed CNOT gates by avoiding the use of bridge qubits. For example, on a line topology the minimum distance to decompose $ZZI$ is 1, but the minimum distance to $ZIZ$ is 3 because we need one CNOT to turn the gadget into $ZZZ$ and then 2 CNOTs to decompose it into $IIZ$. Fig.~\ref{fig:mapping} gives an example of an optimal mapping of a Pauli exponential containing three Pauli gadgets that spread to four qubits restricted to a line topology (qubit connections 0-1-2-3).

\begin{figure}
    \centerline{
    \includegraphics[width=0.20\textwidth]{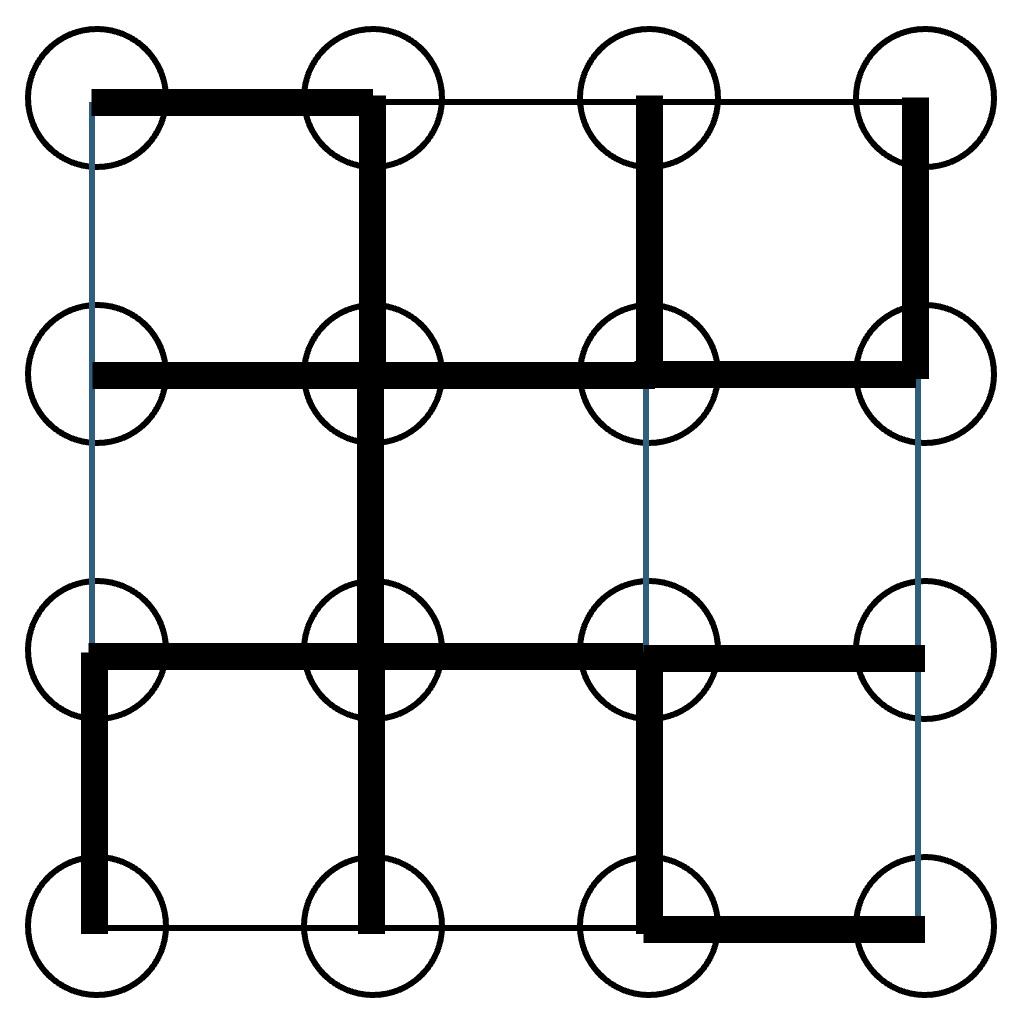}}
    \caption{Example of connectivity tree constructed on 16-grid topology.}
    \label{fig:tree_mapping}
\end{figure}

The mapping algorithm generates an injective mapping from logical qubits to physical qubits and constructs a spanning tree, $G_{tree}$, representing the utilized physical connections (Figure \ref{fig:tree_mapping}). This tree is defined as a subgraph of the hardware topology $G_{topo}$, specifically omitting unused vertices and certain edges to eliminate cycles. The resulting connectivity tree is subsequently utilized by the synthesis algorithm.

The algorithm takes the following inputs:
\begin{enumerate}
\item Pauli Gadgets: A set $\mathcal{G}$, where for each $P \in \mathcal{G}$, $P_i$ denotes the Pauli operator acting on qubit $i$.

\item Coupling Graph: A physical topology $G_{topo} = (V_{topo}, E_{topo})$, where $V_{topo}$ and $E_{topo}$ represent physical qubits and their available coupling constraints, respectively.

\item Logical Qubits: A set $V_{log}$ representing the logical qubits to be mapped.
\end{enumerate}

The outputs consist of a mapping function $f: V_{log} \to V_{topo}$ and the connectivity tree $G_{tree} = (V_{tree}, E_{tree})$, where $V_{tree} \subseteq V_{topo}$ and $E_{tree} \subseteq E_{topo}$. The mapping procedure is defined as follows:

\begin{enumerate}
\item Score all logical qubit-pairs according to the number of Pauli gadgets which have non-$I$ Pauli operators in both qubits.
$$S_{ij} = \sum_{P \in \mathcal{G}} \delta(P_{i}, P_{j}) $$ 

where $\delta(P_i, P_j) = 1$ if $P_i \neq I \land P_j \neq I$ and otherwise 0. 

\item Locate the central physical qubit of the topology: qubit having maximum connectivity and longest minimum distance to qubits having a minimum connectivity. 
$$q^{phys}_{root} = \text{argmax}_{v \in V_{max}} \left( \min_{u \in V_{min}} d(v, u) \right)$$

where $\text{deg}(v)$ is the connectivity (degree) of vertex $v$, $d(u, v)$ is the shortest path distance between verices $u$ and $v$, $V_{max} = \{v \in V \mid \text{deg}(v) = \max_{u \in V} \text{deg}(u)\}$, and $V_{min} = \{v \in V \mid \text{deg}(v) = \min_{u \in V} \text{deg}(u)\}$. Add this physical qubit as a root of the connectivity tree.
$$V_{tree} \gets \{q^{phys}_{root}\}$$

\item Identify logical qubit having most non-$I$ Pauli operators in all Pauli gadgets. 
$$q^{log}_{root} = \text{argmax}_{i \in V_{log}} \sum_{P \in \mathcal{G}} [P_{i} \neq I]$$

Map this logical qubit to the central physical qubit and add the qubits to the sets of mapped qubits.

$$f(q^{log}_{root})=q^{phys}_{root}$$
$$M_{log} \gets \{q^{log}_{root}\}$$
$$M_{physical} \gets \{q^{physical}_{root}\}$$

\item For all non-mapped physical qubits $N_{physical}=V_{topo} \setminus M_{physical}$ which are a neighbor of any mapped physical qubits $M_{physical}$, try all non-mapped logical qubits $N_{log}=V_{log} \setminus M_{log}$: select logical-physical qubit-pair $(q^{log}_n \in N_{log},q^{physical}_n \in N_{physical})$ which have the maximum score between the selected logical qubit $q^{log}_n$ and the logical qubit previously mapped to the neighbour of selected physical qubit $f^{-1}(q^{phys}_m)$. 
$$(q^{log}_n, q^{phys}_n, q^{phys}_m) = \text{argmax}_{(l, q, p) \in \mathcal{C}} S_{l, f^{-1}(p)}$$

Where the set $\mathcal{C}$ is defined as:
\begin{equation*}
\begin{split}
\mathcal{C} = \{ (l, q, p) \mid \,\, & l \in N_{log}, \,\, q \in N_{physical}, \\
& p \in M_{physical}, \,\, (p, q) \in E_{topo} \}
\end{split}
\end{equation*}

Map the selected logical qubit $q_{log}$ to the selected physical qubit $q_{phys}$. 
$$f(q^{log}_n)=q^{phys}_n$$

Add the selected physical qubit $q_{phys}$ to the connectivity tree in addition to an edge between physical qubit-pair $q_{phys}-q_{phys}$. 
$$V_{tree} \gets V_{tree} \cup \{q^{phys}_n\}$$
$$E_{tree} \gets E_{tree} \cup \{(q^{phys}_m, q^{phys}_n)\}$$

Finally, add the qubits to the sets of mapped qubits.
$$M_{log} \gets M_{log} \cup \{q^{log}_n\}$$
$$M_{physical} \gets M_{physical} \cup \{q^{phys}_n\}$$

Continue this loop until all logical qubits are mapped.

\end{enumerate}

The time complexity of the mapping algorithm is polynomial. In the first step, the algorithm analyzes all qubit-pairs. For each qubit-pair, all gadgets need to be processed. So, the time complexity of the first step is $O(gn^2)$ where $n$ is the number of physical qubits and $g$ is the number of gadgets. In the second step, we have to calculate the distance of all physical qubit-pairs, which has a time complexity of $O(n^3)$. The time complexity of the third step is $O(gn)$. 

In the fourth step, every qubit must be mapped. For every qubit, we need to compare the maximum of $n$ non-mapped physical qubits to $n$ non-mapped logical qubits. This has a time complexity of $O(n^3)$. In general, the time complexity of the algorithm is $O(gn^2 + n^3)$.

Our mapping algorithm is somewhat similar to the mapping in the Pauli Forest algorithm presented by Li et al.~\cite{li_pauliforest_2025}. They have the same idea to score which qubits should be placed as a neighbor, but use a different heuristic.

\subsection{Synthesis Algorithm}

During synthesis, the algorithm has two major choices: in which order Pauli gadgets are decomposed, and by which gates each individual Pauli gadget is decomposed. The general idea is to find the shortest path that visits all nodes such that all Pauli gadgets are decomposed. All possible combinations of paths cannot be analyzed because the problem is NP-complete~\cite{schmitz_graph_2023}, so we need a heuristic algorithm.

The simple choice for the problem to choose a Pauli gadget is to greedily always choose the nearest Pauli gadget. This is what the algorithm does. 

The next step to choose the path to decompose the selected gadget is more complex. The path is constructed step by step. Here again, algorithm greedily chooses a step that shortens the distance to the node where the chosen Pauli gadget is decomposed. Usually, there are several such alternatives. From the alternatives, the algorithm chooses one such that on average the distance to all remaining Pauli gadgets is minimized the most. This is the same general idea as in the Pauli Frame graph algorithm~\cite{schmitz_graph_2023, brugiere_faster_2024} (which is not architecture-aware) and in the Parity network synthesis algorithm~\cite{paritynetwork} (which is restricted to phase polynomials rather than Pauli exponentials). 

\begin{figure}
    \centerline{
    \includegraphics[width=0.25\textwidth]{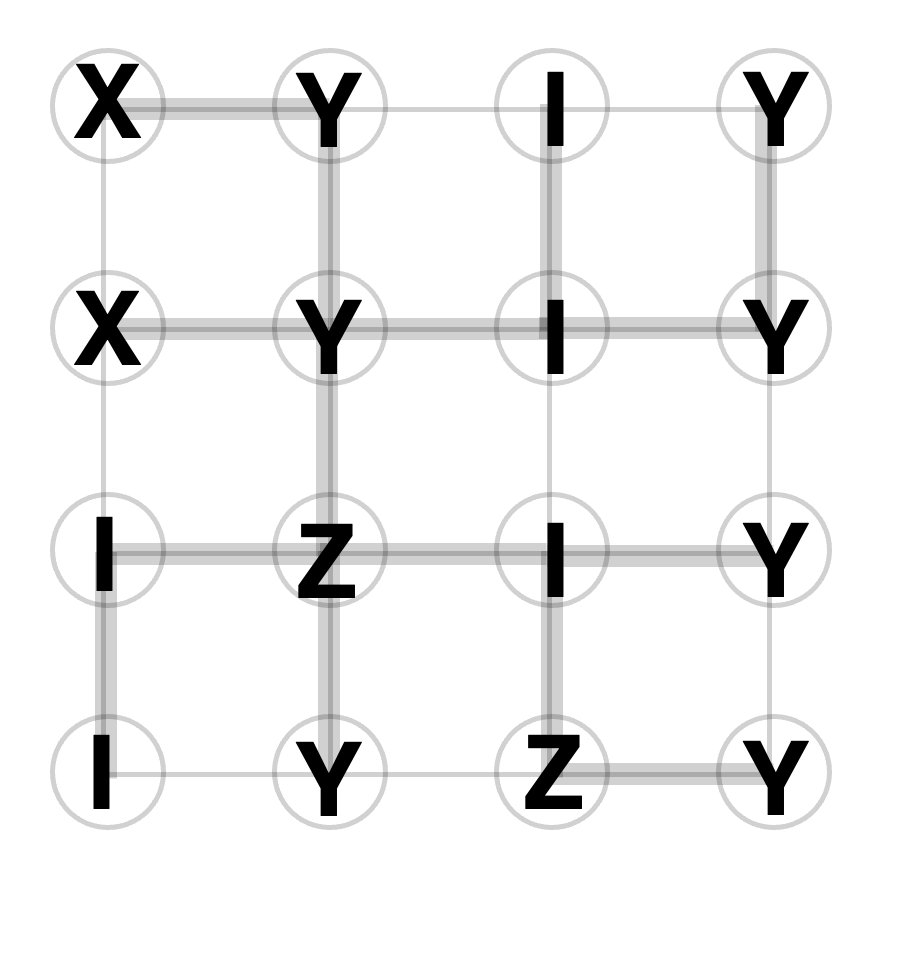}}
    \caption{Example of gadget mapped to tree before pruning.}
    \label{fig:gadget_tree}
\end{figure}

All Pauli gadgets are mapped to the same connectivity tree graph created in the start of the mapping phase (Figure \ref{fig:gadget_tree}). The tree of each gadget is pruned from the original tree so that all leafs of the final tree graph contain non-$I$ Pauli operators. In a single decomposition step, the non-$I$ Pauli operator in one of the leafs of the graph is turned into an $I$ operator by applying a gate set to that qubit ($q_c$) and a qubit adjacent to it ($q_t$). If there exists an $I$ operator on the adjacent qubit (so called bridge qubit), we need first do a step where this qubit is turned into non-$I$ Pauli operator. 

The minimum distance of the gadget can be calculated by adding the number of nodes in the tree graph to the number of nodes containing an $I$ operator. The algorithm updates this tree graph when a gate is commuted through each gadget.

The single decomposing procedure consists of applying single qubit gates to two selected qubits and a CNOT to both qubits (so that the control qubit is $q_c$ and the target qubit $q_t$). The single qubit gate for the control qubit $q_c$ is either the $I$, $V$, or $SV$ gate and the single qubit gate for the target qubit $q_t$ is $V$, $S$, or $VS$. Note that with this choice of surrounding single qubit gates, the direction of the CNOT does not matter since the choice of $SV$ and $VS$ is equivalent to that.

So, there are nine different gate combinations in total for each step. This is half of those that Rustiq uses~\cite{brugiere_faster_2024}, and the same as Schmitz et al.~\cite{schmitz_graph_2023}. All these combinations have a unique effect on each possible Pauli operator combinations when the gates are commuted through Pauli gadgets in terms of how they convert the Pauli letters in the Pauli operator to the $I$ operator and vise versa. Other gate combinations would not produce any unique effect compared to these.

When choosing the next step in the decomposition process, the algorithm compares all possible choices of $q_c$ (so leafs of the current tree graph of the selected gadget). For each of these, the algorithm selects from the nine gate combinations exactly those combinations that turn $q_c$ to $I$. The algorithm then calculates which effect of these gate combinations has to the minimum distance of remaining gadgets. The algorithm then selects the qubit-pair and gate combination that brings the remaining gadgets closest on average.

The time-complexity of the algorithm is polynomial in the number of qubits $n$ and the number of gadgets $g$. The algorithm must process every gadget. For every gadget, there are a maximum of $2n-2$ steps to remove the leaves and process any possible bridge qubits. For every step, there are a maximum of $n-1$ choices for $q_c$ since the root of the qubit tree will never be a leaf, or else the gadget is trivially decomposable. 

For each of these choices, the algorithm analyzes the effect of all nine gate combinations on all (remaining) gadgets. Lastly, the algorithm updates the graphs of all remaining gadgets. Updating a single graph has a time complexity of $O(1)$ as this will add or remove a single edge. So, the time complexity is $g(2n-2)9(n-1)g = 18g^2n^2+18 \simeq O(g^2n^2)$.  

\begin{figure}
    \centerline{
    \includegraphics[width=0.45\textwidth]{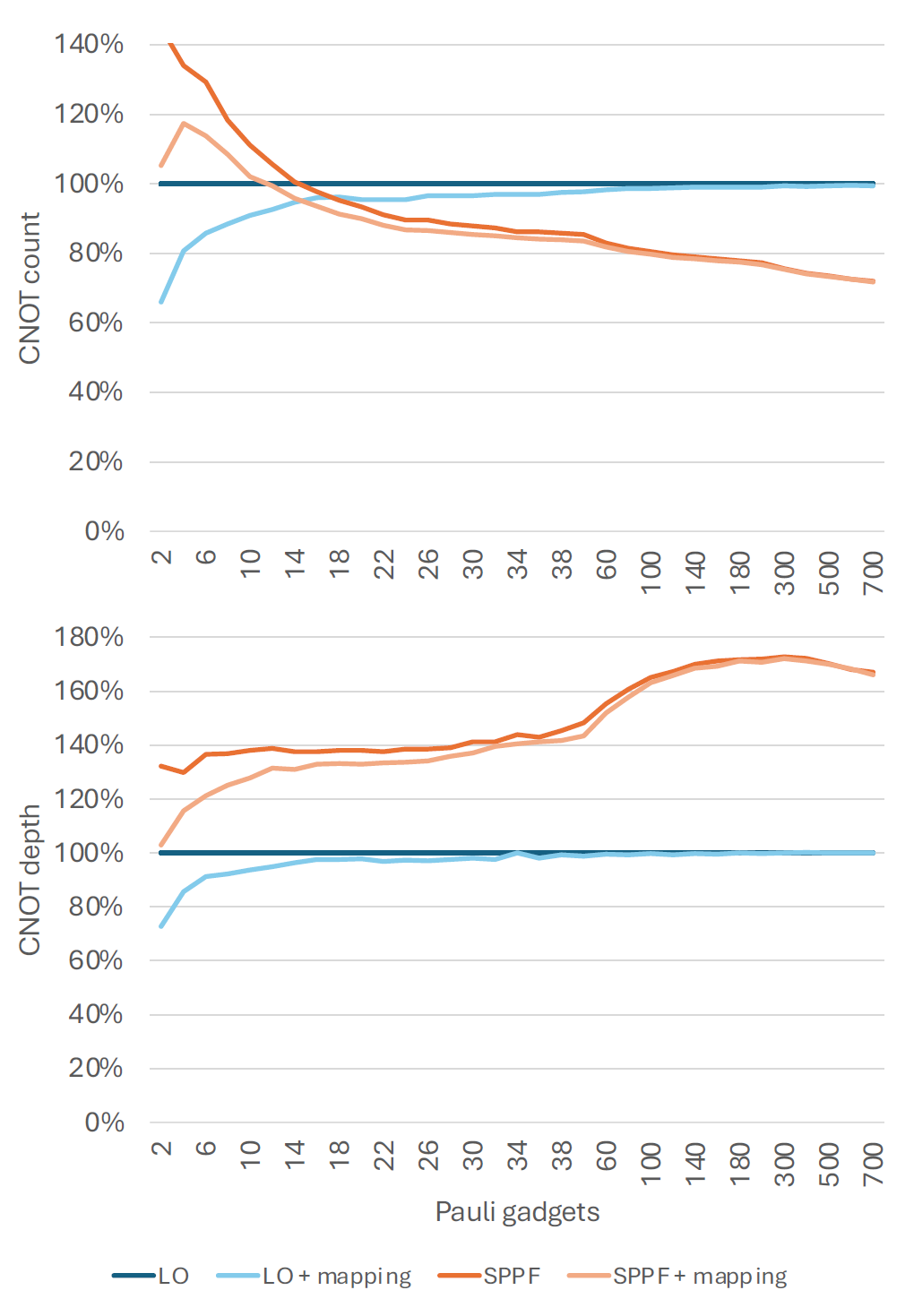}}
    \caption{Results on line topology with 16 qubits. Lexicographical ordering (LO) without mapping is the baseline.}
    \label{fig:line16}
\end{figure}

\section{Results}

The novel mapping algorithm was compared to the algorithm introduced in Huang et al.~\cite{huang_redefining_2024} and those used in Qiskit and TKET. The algorithm is implemented was a part of Pauliopt library and is available on Github\footnote{https://github.com/vuorenkoski/pauliopt/tree/main/experiments}. All experiments can be replicated using that repository. The results presented here were produced using a MacBook Air with M3 CPU and 16GB memory. 

We compared to the algorithm presented in Huang et al.~\cite{huang_redefining_2024} (which is referred to as lexicographical ordering algorithm) to our proposed algorithm using three topologies: line, grid and Guadalupe, each having 16 qubits. 

The performance was measured with Pauli exponentials ranging from $2-700$ random gadgets covering $16$ qubits. Each gadget contained uniformly at random $1-16$ non-$I$ Pauli operators at random positions. The Pauli operator for the non-$I$ positions was uniformly sampled from $X$, $Y$ and $Z$. The gadgets' angles were uniformly sampled from $\pi$, $\frac{\pi}{2}$, $\frac{\pi}{4}$, $\frac{\pi}{8}$, and $\frac{\pi}{16}$. For each size, $200$ random Pauli exponentials were sampled, and the results are averages of those samples. All samples were synthesized with four algorithms: lexicographical ordering with random mapping (LO), lexicographical ordering with our mapping algorithm (LO + mapping), our synthesis algorithm without mapping (SPPF), and with mapping (SPPF + mapping).

When the target topology is a line of $16$ qubits, our algorithm produced fewer CNOT gates than the Lexicographical ordering algorithm when there were more than 14 gadgets (Fig.~\ref{fig:line16}). The difference increased when the number of gadgets increased, being $72\%$ when the number of gadgets was $700$. The results are the opposite in terms of the CNOT depth: the lexicographical ordering algorithm consistently obtains a lower depth. 

The mapping algorithm decreased the number of CNOT gates and the CNOT depth compared to random mapping, but only when the number of gadgets was approximately $<40$. The results were similar between Lexicographical ordering and our synthesis algorithm.

\begin{figure}
    \centerline{
    \includegraphics[width=0.45\textwidth]{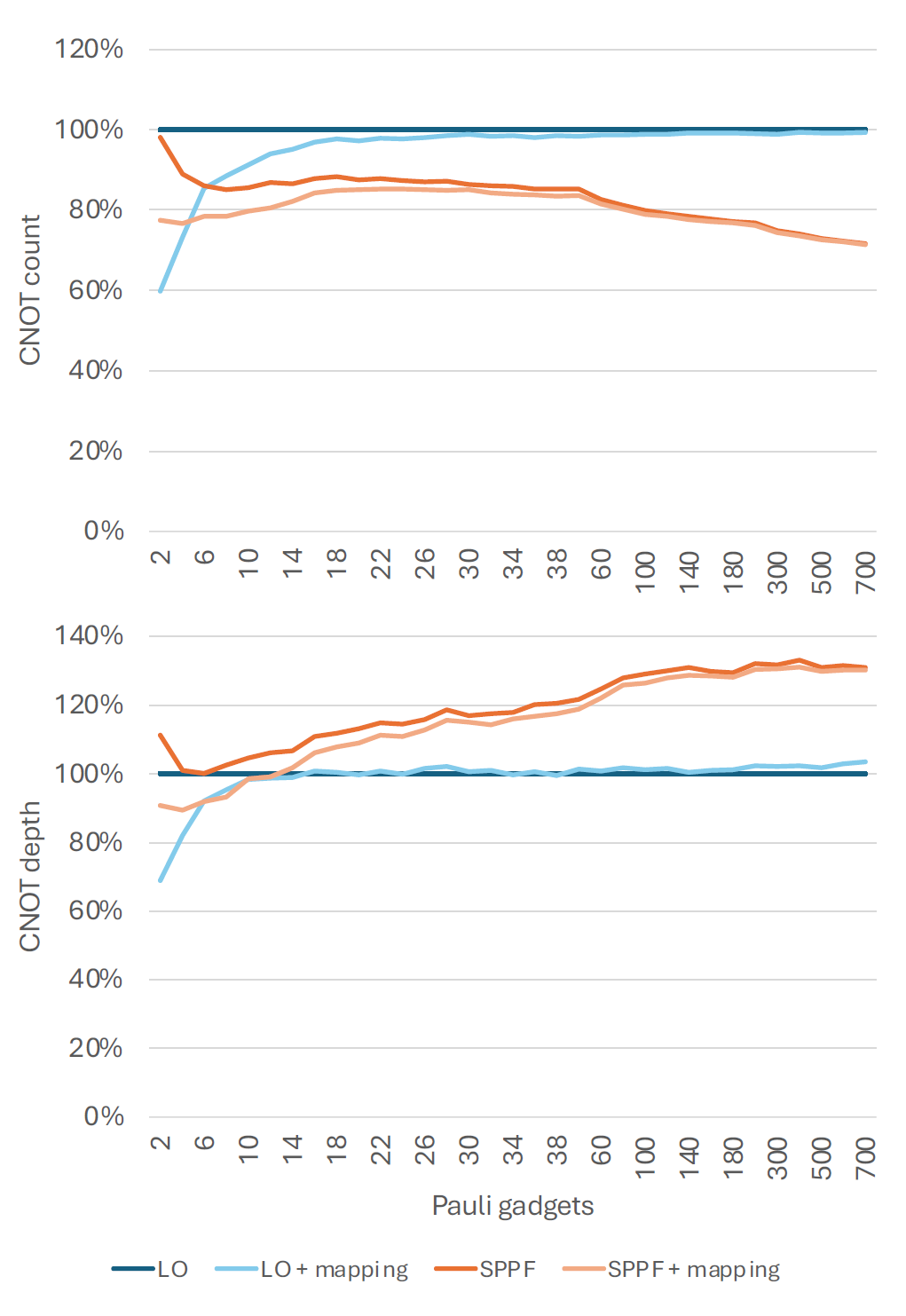}}
    \caption{Results on Guadalupe topology with 16 qubits. Lexicographical ordering (LO) without mapping is the baseline.}
    \label{fig:guadalupe}
\end{figure}

When the Pauli exponentials were mapped to the Guadalupe topology, the results were similar. In terms of CNOT count, our algorithm performed better even for small gadgets (Fig.~\ref{fig:guadalupe}). The difference increased as the number of gadgets increased with $71\%$ when the number of gadgets was $700$. The difference in CNOT depth was not as large as in the line topology. The maximum observed difference is $31\%$.

\begin{figure}
    \centerline{
    \includegraphics[width=0.45\textwidth]{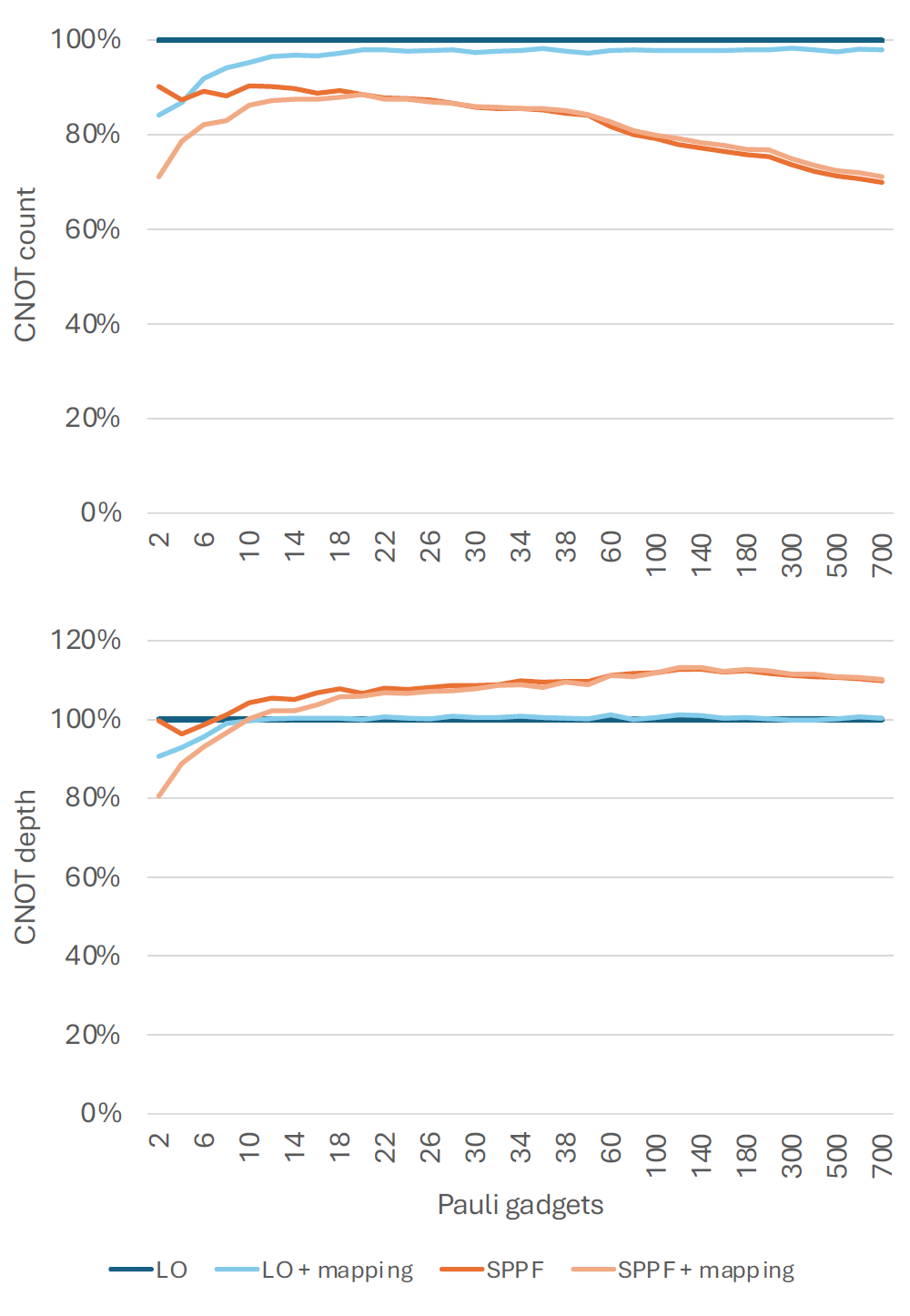}}
    \caption{Results on grid topology with 16 qubits. Lexicographical ordering (LO) without mapping is the baseline.}
    \label{fig:grid16}
\end{figure}

The results on Pauli exponentials mapped to the grid topology were very similar to those for the Guadalupe topology (Fig.~\ref{fig:grid16}). The most notable difference was that the maximum observed difference of the CNOT depth was only $13\%$.

\begin{figure}
    \centerline{
    \includegraphics[width=0.45\textwidth]{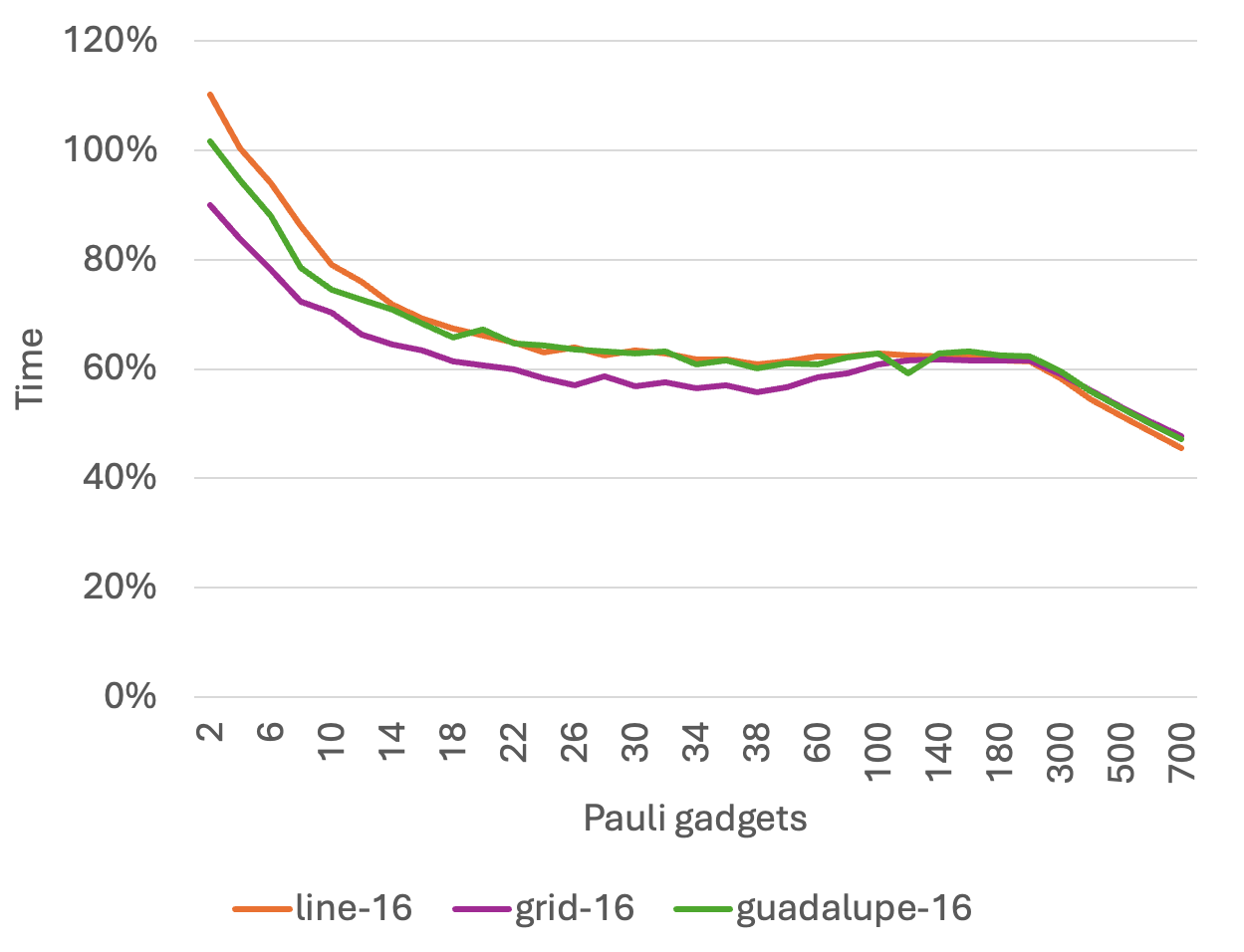}}
    \caption{Time taken by algorithms in different topologies. Lexicographical ordering (LO) without mapping is the baseline.}
    \label{fig:time}
\end{figure}

Our algorithm also performed better in terms of runtime (Fig.~\ref{fig:time}). For Pauli exponentials with more than $4$ gadgets, our algorithm was faster. The time difference increased as the number of gadgets increased, up to about $50\%$ when the number of gadgets was $700$ on all topologies. 

The algorithm was also compared with the decomposition of random Pauli exponentials using Qiskit (version 2.1.1) and TKET (version 2.9.1). In this experiment, we used the $127$ qubit Brisbane topology. We used varying sizes of Pauli exponentials ($10-100$ gadgets). Each gadget contained uniformly randomly $1-16$ non-$I$ Pauli operators with random positions. The Pauli operator for non-$I$ positions was uniformly sampled from $X$, $Y$ and $Z$. The gadgets' angles were uniformly sampled from $\pi$, $\frac{\pi}{2}$, $\frac{\pi}{4}$, $\frac{\pi}{8}$, $\frac{\pi}{16}$. For each size, $20$ random Pauli exponentials were sampled and the results were averaged.

In Qiskit, we used two different strategies for synthesizing the circuit from an evolution gate: default synthesis and the Rustiq algorithm~\cite{brugiere_faster_2024}. In both cases, the resulting circuit was transpiled to the target topology using SABRE~\cite{li_tackling_2019} as layout method and routing method with optimization level 3. In TKET, we used the algorithm adapted from Schmitz et al.~\cite{schmitz_graph_2023}, and the \emph{CXMappingPass} was used to transpile the resulting circuit to the target topology, after which we used \emph{FullPeepholeOptimise} step to further optimize the circuit.

\begin{figure}
    \centerline{
    \includegraphics[width=0.44\textwidth]{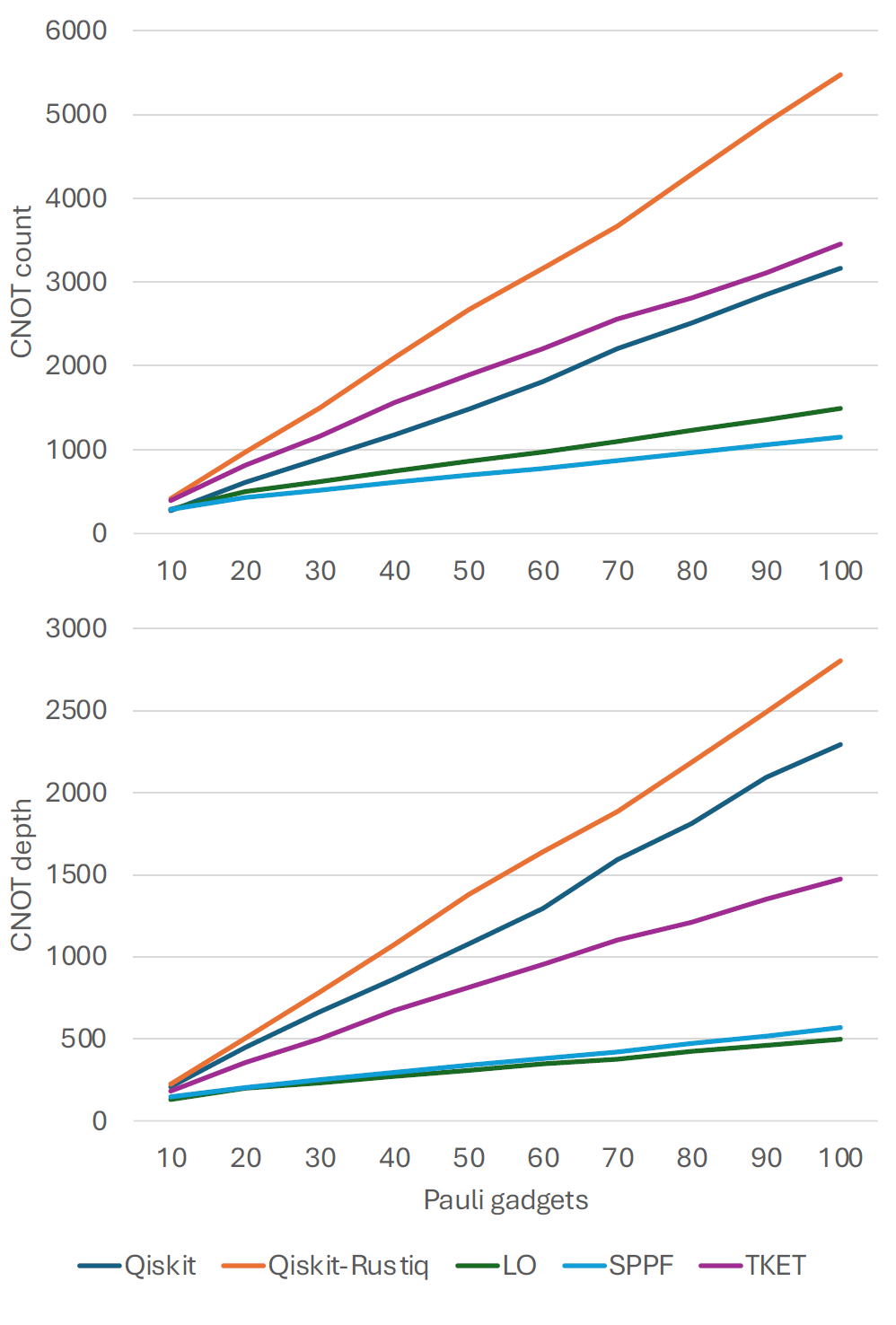}}
    \caption{Results on 16 qubit Pauli exponentials mapped to 127 qubit Brisbane topology.}
    \label{fig:qiskit}
\end{figure}

The results show that our algorithm and Lexicographical ordering algorithm were significantly better than the other algorithms, both in terms of CNOT count and CNOT depth (Figure~\ref{fig:qiskit}). Much smaller differences can be observed between our algorithm and Lexicographical ordering algorithm. In $100$ gadget exponentials, our algorithm produced $29\%$ fewer CNOTs but the depth was $13\%$ larger than what Lexicographical ordering algorithm produced. 

Additionally, we compared our algorithm to Lexicographical ordering algorithm with a selection of Pauli exponentials representing UCCSD molecular ans\"atze mapped to IBM topologies Quito, Guadalupe, Nairobi and Mumbai (Table~\ref{tab:molecules}). The Pauli exponentials had different numbers of qubits and gadgets. The Pauli exponentials were mapped to the smallest topology as possible. Here, our algorithm performed better in terms of CNOT count, CNOT depth, and time. Our algorithm produced $29\%-68\%$ fewer CNOTs than the Lexicographical ordering algorithm. It also produced circuits having smaller CNOT depth except for one exponential (a reduction between $11\%-65\%$). The differences of runtime were significant. The runtime of our algorithm was $7\%-91\%$ faster than that of lexicographical ordering (with the exception of the smallest molecule). 

\begin{table*}[htb]
\caption{Results on UCCSD molecular ans\"atze.}
\label{tab:molecules}
\begin{center}
\begin{tabular}{|c|c|c|c|c|c|c|c|c|c|c|}
\hline
\multicolumn{4}{|c|}{\textbf{}} &\multicolumn{3}{|c|}{\textbf{CNOT count}} &\multicolumn{3}{|c|}{\textbf{CNOT depth}}&\textbf{}\\

\cline{5-10} 
\textbf{Ansatzs} & \textbf{Backend}& \textbf{Qubits}& \textbf{\textit{Gadgets}} &
\textbf{SPPF} & \textbf{LO}& \textbf{diff \%}&
\textbf{SPPF} & \textbf{LO}& \textbf{diff \%}&
\textbf{time diff. \%}\\
\hline

H2\_BK\_sto3g & quito & 4 & 12 & 22 & 35 & -37.1 & 22 & 35 & -37.1 & 5.2 \\
H2\_JW\_sto3g & quito & 4 & 12 & 25 & 38 & -34.2 & 25 & 38 & -34.2 & -7.0 \\
H2\_P\_631g & nairobi & 6 & 158 & 217 & 433 & -49.9 & 166 & 332 & -50.0 & -50.6 \\
H4\_P\_sto3g & nairobi & 6 & 164 & 208 & 471 & -55.8 & 170 & 344 & -50.6 & -52.0 \\
H2\_BK\_631g & guadalupe & 8 & 84 & 236 & 443 & -46.7 & 172 & 301 & -42.9 & -44.3 \\
H2\_JW\_631g & guadalupe & 8 & 84 & 229 & 414 & -44.7 & 169 & 272 & -37.9 & -43.2 \\
H4\_BK\_sto3g & guadalupe & 8 & 160 & 382 & 878 & -56.5 & 280 & 523 & -46.5 & -55.9 \\
H4\_JW\_sto3g & guadalupe & 8 & 160 & 410 & 883 & -53.6 & 290 & 552 & -47.5 & -55.0 \\
LiH\_P\_sto3g & guadalupe & 10 & 630 & 1279 & 3472 & -63.2 & 840 & 1933 & -56.5 & -81.0 \\
NH\_P\_sto3g & guadalupe & 10 & 630 & 1279 & 3472 & -63.2 & 840 & 1933 & -56.5 & -80.2 \\
BeH2\_P\_sto3g & guadalupe & 12 & 1085 & 2891 & 8482 & -65.9 & 1810 & 4372 & -58.6 & -84.2 \\
CH2\_P\_sto3g & guadalupe & 12 & 2109 & 4583 & 14909 & -69.3 & 2686 & 7740 & -65.3 & -91.1 \\
H2O\_P\_sto3g & guadalupe & 12 & 1085 & 2699 & 8077 & -66.6 & 1682 & 4015 & -58.1 & -86.8 \\
LiH\_BK\_sto3g & guadalupe & 12 & 640 & 2665 & 4250 & -37.3 & 1601 & 2261 & -29.2 & -62.1 \\
LiH\_JW\_sto3g & guadalupe & 12 & 640 & 1910 & 4331 & -55.9 & 1147 & 2385 & -51.9 & -70.1 \\
NH\_BK\_sto3g & guadalupe & 12 & 640 & 2081 & 4788 & -56.5 & 1315 & 2433 & -46.0 & -76.5 \\
NH\_JW\_sto3g & guadalupe & 12 & 640 & 1979 & 4884 & -59.5 & 1234 & 2608 & -52.7 & -73.7 \\
BeH2\_BK\_sto3g & guadalupe & 14 & 1488 & 6675 & 12419 & -46.3 & 4009 & 5628 & -28.8 & -78.8 \\
BeH2\_JW\_sto3g & guadalupe & 14 & 1488 & 5183 & 12181 & -57.5 & 2930 & 5908 & -50.4 & -81.4 \\
CH2\_BK\_sto3g & guadalupe & 14 & 1488 & 5778 & 12919 & -55.3 & 3331 & 6182 & -46.1 & -81.4 \\
CH2\_JW\_sto3g & guadalupe & 14 & 1488 & 4873 & 13433 & -63.7 & 2808 & 6398 & -56.1 & -82.2 \\
H2O\_BK\_sto3g & guadalupe & 14 & 1000 & 3850 & 8675 & -55.6 & 2205 & 3877 & -43.1 & -75.5 \\
H2O\_JW\_sto3g & guadalupe & 14 & 1000 & 3637 & 7967 & -54.4 & 2074 & 3683 & -43.7 & -79.3 \\
H4\_P\_631g & guadalupe & 14 & 2912 & 7875 & 24224 & -67.5 & 4318 & 10984 & -60.7 & -92.7 \\
H4\_BK\_631g & mumbai & 16 & 1440 & 7085 & 13852 & -48.9 & 4093 & 4585 & -10.7 & -83.7 \\
H4\_JW\_631g & mumbai & 16 & 1440 & 5446 & 11727 & -53.6 & 3059 & 5267 & -41.9 & -82.9 \\
H8\_BK\_sto3g & mumbai & 16 & 2688 & 10710 & 28018 & -61.8 & 6104 & 9355 & -34.8 & -88.6 \\
H8\_JW\_sto3g & mumbai & 16 & 2688 & 9399 & 29280 & -67.9 & 5383 & 9118 & -41.0 & -89.8 \\
NH3\_BK\_sto3g & mumbai & 16 & 2340 & 8931 & 25280 & -64.7 & 5173 & 8400 & -38.4 & -88.8 \\
NH3\_JW\_sto3g & mumbai & 16 & 2340 & 8486 & 23182 & -63.4 & 4664 & 7937 & -41.2 & -87.7 \\
C2\_P\_sto3g & mumbai & 18 & 2950 & 12138 & 37786 & -67.9 & 6649 & 11144 & -40.3 & -93.1 \\
N2\_P\_sto3g & mumbai & 18 & 2950 & 12111 & 35602 & -66.0 & 6455 & 10935 & -41.0 & -91.5 \\
HCl\_BK\_sto3g & mumbai & 20 & 684 & 2910 & 6308 & -53.9 & 1643 & 2712 & -39.4 & -75.8 \\
HCl\_JW\_sto3g & mumbai & 20 & 684 & 3365 & 5807 & -42.1 & 1897 & 2677 & -29.1 & -70.1 \\
LiH\_BK\_631g & mumbai & 22 & 3240 & 25480 & 35792 & -28.8 & 12954 & 11586 & 11.8 & -82.6 \\
LiH\_JW\_631g & mumbai & 22 & 3240 & 14646 & 33673 & -56.5 & 7816 & 11058 & -29.3 & -89.3 \\

\hline
\end{tabular}
\end{center}
\end{table*}

\section{Discussion}

Our algorithm has similarities with several previous algorithms: mapping resembles that of Pauli Forest~\cite{li_pauliforest_2025}, the idea of presenting synthesis algorithm as a shortest path problem was used in the Pauli Frame graph algorithm~\cite{schmitz_graph_2023}, Clifford tableau synthesis is used in algorithm designed by Huang et al.~\cite{huang_redefining_2024}. The novel contribution of our work is to bring these ideas together to architecture-aware synthesis.

In general, our algorithm proved to be better in terms of CNOT count compared to selected existing algorithms. It was efficient especially when decomposing Pauli exponentials constructed from molecular ans\"atze used in Hamiltonian simulation. Our algorithm also performed better in terms of runtime compared to the algorithm developed by Huang et al.~\cite{huang_redefining_2024}.

We made a few very interesting observations. One is that the CNOT depth of the proposed algorithm is worse than the algorithm developed by Huang et al.~\cite{huang_redefining_2024} in random Pauli exponentials, but not for ans\"atze of real molecules. Molecular ans\"atze seem to have some special features that are beneficial to our algorithm. Indeed, the molecular ans\"atze used in our study are not uniformly random as the random Pauli exponential sample. One direction of future research is to find out what the significant difference is, and how to utilize that difference. 

The initial mapping procedure in our proposed algorithm performed well only in Pauli exponentials with a small number of gadgets (below around $20$ gadgets in our experiments). When the size of Pauli exponential is larger, the mapping algorithm performs similarly to random mapping. This is expected because when there are more gadgets, more qubits will be interacting with each other, and a poor initial mapping becomes less detrimental to the synthesis process.

Namely, the mapping algorithm is naturally better than random mapping in rather trivial cases where only part of the qubits of sparsely connected topology are used in the exponential. In these cases, the random mapping would disperse the used qubits throughout the topology.

\bibliographystyle{IEEEtran}
\bibliography{IEEEabrv,references}

\end{document}